\documentclass[preprint,sort&compress]{elsarticle}

\usepackage[utf8]{inputenc}
\usepackage{amsmath}
\usepackage{amsfonts}
\usepackage{amssymb}
\usepackage{color}
\usepackage{graphicx}
\usepackage{bbold}
\usepackage{mathtools}
\usepackage{fancyhdr}
\usepackage{multirow}
\usepackage{tikzsymbols}

\usepackage{pifont}
\newcommand{\cmark}{{ \ding{51}}}
\newcommand{\xmark}{{\ding{55}}} 

\title{Fibonacci Fast Convergence for Neutrino Oscillations in Matter}

\usepackage[colorlinks]{hyperref}
\makeatletter
\hypersetup{allcolors=[rgb]{0,0,1},pdftitle=\@title}
\makeatother

\newcommand{\eps}{\epsilon}
\newcommand{\orcid}[1]{orcid: \href{https://orcid.org/#1}{#1}}

\begin{document}

\begin{frontmatter}

\author[bnl]{Peter B.~Denton\fnref{fn2}}
\ead{pdenton@bnl.gov}
\fntext[fn2]{\orcid{0000-0002-5209-872X}}
\address[bnl]{Physics Department, Brookhaven National Laboratory, Upton, New York 11973, USA}

\author[fnal]{Stephen J.~Parke\fnref{fn3}}
\ead{parke@fnal.gov}
\fntext[fn3]{\orcid{0000-0003-2028-6782}}
\address[fnal]{Theoretical Physics Dept., Fermi National Accelerator Laboratory, Batavia, IL 60510, USA}

\author[uc]{Xining Zhang\fnref{fn1}}
\ead{xining@uchicago.edu}
\fntext[fn1]{\orcid{0000-0001-8959-8405}}
\address[uc]{Enrico Fermi Institute \& Dept.~of Physics, University of Chicago, Chicago, IL 60637, USA}

\begin{abstract}
Understanding neutrino oscillations in matter requires a non-trivial diagonalization of the Hamiltonian.
As the exact solution is very complicated, many approximation schemes have been pursued.
Here we show that one scheme, systematically applying rotations to change to a better basis, converges exponentially fast wherein the rate of convergence follows the Fibonacci sequence.
We find that the convergence rate of this procedure depends very sensitively on the initial choices of the rotations as well as the mechanism of selecting the pivots.
We then apply this scheme for neutrino oscillations in matter and discover that the optimal convergence rate is found using the following simple strategy: first apply the vacuum (2-3) rotation and then use the largest off-diagonal element as the pivot for each of the following rotations.
The Fibonacci convergence rate presented here may be extendable to systems beyond neutrino oscillations.
\end{abstract}

\end{frontmatter}

\fancyhf{}
\renewcommand{\headrulewidth}{0pt}
\setlength{\headheight}{28pt}
\rhead{FERMILAB-PUB-19-462-T}
\rfoot{\footnotesize \today}
\thispagestyle{fancy}

\section{Introduction}
Measurements of neutrino oscillations have triggered an immense interest in gaining a better understanding of neutrino oscillations, specifically in the presence of the Wolfenstein matter effect \cite{Wolfenstein:1977ue}.
Due to the complexity of the exact analytic solution for more than two flavors and the presence of the $\cos(\frac13\cos^{-1}(\cdots))$ term, the exact expressions are no more insightful than numerically diagonalizing the Hamiltonian directly.
To address this, many approximate formulas have been developed, see Ref.~\cite{Parke:2019vbs} for a 2019 review.
A recent example is a rotation method known as the Jacobi method \cite{jacobi} which has been applied to neutrino oscillations to calculate the energy eigenvalues and eigenstates to high precision with simple structure and high calculation speed \cite{Petcov:1986qg,Honda:2006hp,Honda:2006gv,Kopp:2006wp,Agarwalla:2013tza,Blennow:2013rca,Minakata:2015gra,Denton:2016wmg,Denton:2018hal,Denton:2018fex,Parke:2019vbs,Martinez-Soler:2019nhb,Parke:2019jyu,Yue:2019qat}.
The principle of this method is performing rotations to resolve the crossings of the diagonal elements and to reduce the size of the off-diagonal elements of the effective Hamiltonian.

In this paper we further expand upon the properties of the rotation method.
We report a phenomenon that in the context of three neutrino oscillations wherein precision of the approximation will be improved very rapidly with the number of rotations implemented provided that the sequence of rotations is chosen carefully.
We identify this sequence and show that the order of the uncertainties follows the Fibonacci series and thus grows exponentially.
This feature provides a fast way of high precision calculation of neutrino oscillations in matter. 

The structure of this paper is listed following.
In section \ref{section:derivation} we review the effects of rotations on the crossings and sizes of the off diagonal elements in the Hamiltonian.
Section \ref{section:numerical} applies this method to three flavor neutrino oscillation in matter paying particular attention to the size of the corrections.
We also identify the sequence of oscillation that leads to Fibonacci convergence in the size of the corrections and show numerically how well it works in the context of neutrino oscillations in matter.
Section \ref{section:conclusion} is the conclusion and summary of this paper's contents.
Other materials we believe necessary can be found in the Appendices.

\section{Derivation of the Main Result} \label{section:derivation}
\subsection{Preliminary rotations}
The three neutrino problem in matter requires solving a fully populated complex $3\times 3$ Hermitian matrix for its eigenvectors and eigenvalues.
Given those (or just the eigenvalues, see \cite{Yokomakura:2000sv,Kimura:2002hb,Kimura:2002wd,Denton:2019ovn,Denton:2019pka}) determining the oscillation probabilities is straightforward.
Because of the matter effect, the PMNS matrix with the vacuum parameters no longer diagonalizes the Hamiltonian and the eigenvalues are also altered.
The Hamiltonian is typically split into the large diagonal elements and the smaller off diagonal elements. Applying perturbation theory at this point suffers from two problems: 1) the zeroth order eigenvalues (diagonal elements) cross at two matter potential values and 2) the perturbative terms (off-diagonal elements) are not particularly small.
Recently a rotation method has been applied to overcome the above defects \cite{Agarwalla:2013tza,Denton:2016wmg,Denton:2018fex}.
The rotations applied can be used to address both issues: they can eliminate the largest off-diagonal elements in the perturbing Hamiltonian and they cause the level crossings to repel each other.

For an arbitrary $n\times n$ Hermitian matrix $\mathcal{H}$, we choose two diagonal elements $\mathcal{H}_{pp}$ and $\mathcal{H}_{qq}$ and the two corresponding off-diagonal element $\mathcal{H}_{pq}$ and $\mathcal{H}_{qp}$. The selected four elements form a $2\times 2$ Hermitian submatrix $h$,
\begin{equation}
\mathrm{h}=\begin{pmatrix}
\mathcal{H}_{pp}&\mathcal{H}_{pq}\\
\mathcal{H}_{qp}&\mathcal{H}_{qq}
\end{pmatrix}\,.
\label{eq:2x2submatrix}
\end{equation}
It can be diagonalized by a single $2\times2$ complex rotation
\begin{equation}
\mathrm{u}=
\begin{pmatrix}
\cos\alpha  & e^{i\beta}\sin\alpha  \\
-e^{-i\beta}\sin\alpha  & \cos\alpha
\end{pmatrix}\,,
\label{eq:u}
\end{equation}
namely
\begin{equation}
\mathrm{u}^\dagger\,\mathrm{h}\,\mathrm{u}=\text{diag}(\lambda_u,\,\lambda_v)\,,
\end{equation}
where
\begin{equation}
\alpha=\frac{1}{2}\arctan\frac{2|\mathcal{H}_{pq}|}{\mathcal{H}_{qq}-\mathcal{H}_{pp}}\,,\quad
\beta=\text{Arg}[\mathcal{H}_{pq}]\,,
\label{eq:rotationangle}
\end{equation}
and the new diagonal elements are
\begin{equation}
\lambda_{u,v}=\frac{1}{2}\left[\mathcal{H}_{pp}+\mathcal{H}_{qq}\mp \sqrt{(\mathcal{H}_{pp}-\mathcal{H}_{qq})^2+4|\mathcal{H}_{pq}|^2}\right]\,.
\label{eq:lambdapm}
\end{equation}
From Eq.~\ref{eq:lambdapm} we see that the gap between $\lambda_{u,v}$ is at least $2|\mathcal{H}_{pq}|$ so any level crossings of the chosen diagonal elements will be resolved.

In a larger matrix the question of how to identify the correct pivot to use when selecting the relevant $2\times2$ submatrix for Eq.~\ref{eq:2x2submatrix} is a nontrivial one and is the central point of this paper.
While there are a number of strategies present in the context of computer science, for our case there are two obvious,  simple strategies:
\begin{itemize}
\item pick the largest (in absolute value) off-diagonal element (LODE),\\  largest  $|\mathcal{H}_{pq}|$ from above,\\[2mm]
or
\item pick the term that results in the largest (in absolute value) rotation (LROT), largest $|\alpha|$, Eq.~\ref{eq:rotationangle}.
\end{itemize}
In many cases these strategies are equivalent, but they can produce different results when two diagonal elements are close together or cross, such as near the solar resonance.
We require that whatever strategy we use is precise for all neutrino and anti-neutrino energies.

After exploring many cases, we find that, within the context of neutrino oscillations, the two strategies are equivalent in most cases, except near the solar resonance where a LODE and LROT prefer a different rotation pivot in a key early step.
The LROT strategy performs considerably worse, as we show later.

Since the rotation will not increase the scales of any other elements, if the chosen $\mathcal{H}_{pq}$, the pivot, is the LODE element of the full matrix, the leading scale of perturbative terms (off-diagonal elements) has been reduced. 
The above process can be repeated.
By implementing a series of rotations one can eliminate all the crossings of the eigenvalues and squeeze the off-diagonal elements as much as desired.

This procedure, selecting the LODE element, maximizes the precision of the entire matrix.
If however, only certain elements of the matrix are necessary for a given calculation, different techniques may be more optimal.
In the context of neutrino oscillations, our goal is to provide as unified of a framework as possible to apply equally to all channels and all energies.

\subsection{Fibonacci recursive process}

The perturbative method used in \cite{Denton:2016wmg} was compared to using additional rotations in \cite{Denton:2018fex} i.e.~using more rotations to further reduce the off-diagonal elements to enhance the precision without using perturbation theory.
It has been shown that successive rotations will initially match the precision of perturbation theory. 
In this subsection we show that implementing additional rotations will be more efficient to achieve very high order precision than perturbative expressions, after a sufficient number of rotations.

For simplicity we focus on a $3\times 3$ Hamiltonian, although our results may be able to be generalized to other more complicated cases.
After some number of rotations, the Hamiltonian is
\begin{equation}
H=H_0+H_1\,,
\end{equation}
where $H_0$ is a diagonal Hamiltonian and $H_1$ is the perturbative part where all diagonal elements vanish.
That is,
\begin{equation}
H_0=
\begin{pmatrix}
\lambda_1&  & \\
& \lambda_2 & \\
&  &\lambda_3
\end{pmatrix}
\end{equation}
and 
\begin{equation}
H_1=
\begin{pmatrix}
&  \epsilon^a\,x   &\\
\epsilon^a\,x^*& & \epsilon^b\,y\\
&\epsilon^b\,y^* &
\end{pmatrix}\,,
\end{equation}
where $0<\epsilon\ll1$ is a small scale, $a,b$ are some positive numbers, and the matrix is scaled such that $|x|,\,|y|\sim \mathcal{O}(1)$.
Here we have assumed that $(H_1)_{13}=0$, but 
depending on where we are in the sequence of rotations, the pair of vanishing off-diagonal elements of $H_1$ could also be  $(H_1)_{12}$ or  
$(H_1)_{23}$, it will not affect the following deviation.

Next, we assume that $b\geq a>0$, thus  $(H_1)_{12}=\epsilon^a\,x$ is the leading order off-diagonal element now so we should implement a rotation in the (1-2) sector assuming we are implementing the LODE strategy and assuming $(\lambda_2-\lambda_1)\sim{\cal O}(1)$.  Substitute $H_1$ into Eq.~\ref{eq:rotationangle} we get 
\begin{equation}
\alpha_{12}=\frac{1}{2}\arctan\frac{2\epsilon^a\,|x|}{(\lambda_2-\lambda_1)}\,,\quad
\beta_{12}=\text{Arg}[x]\,. \label{eq:alpha beta13}
\end{equation}
After this rotation the perturbative Hamiltonian in the new basis becomes
\begin{equation}
H_1^\prime= \epsilon^b
\begin{pmatrix}
&&-y\sin\alpha_{12}\,e^{i\beta_{12}}  \\
 &  &y\cos\alpha_{12}\\
-y^*\sin\alpha_{12}\,e^{-i\beta_{12}} & y^*\cos\alpha_{12} &
\end{pmatrix}\,.
\end{equation}
By Eq.~\ref{eq:alpha beta13} we see that $\sin\alpha_{12}\sim\mathcal{O}(\epsilon^a)$, therefore the orders the $H_1^\prime$'s elements are $(H_1^\prime)_{23}\sim\mathcal{O}(\epsilon^{b})$ and $(H_1^\prime)_{13}\sim\mathcal{O}(\epsilon^{a+b})$.
Then $H_1^\prime$ can take the place of $H_1$ and one more rotation in the (2-3) sector will extinguish the element $(H_1^\prime)_{23}$.

The above argument however can fail if $(\lambda_2-\lambda_1)$ is small, unless the corresponding $y$ is also smaller than expected.
The smallest $(\lambda_2-\lambda_1)$ can be is $2\eps^a|x|$ exactly on resonance, as shown in Eq.~\ref{eq:lambdapm}.
In this case there is no suppression from $\eps$, but there is still is a slight suppression; for $|x|=1$, we find that $\sin\alpha_{12}=\frac12\sqrt{2-\sqrt{2}}\approx0.38$.
So even in the worst possible case, there is still improvement from the rotation, although it is slow.
We will show in the  next section that at the solar resonance, where $(\lambda_2-\lambda_1)\sim \epsilon$, there is a sequence of rotation such that  the corresponding $y \sim s_{13} \epsilon$.
Then for the corrections to the eigenvectors, there is a cancellation between the numerator and denominator, thus saving the reduction  in the size of the corrections.  
We emphasize that this cancellation occurs only for a special choice of the sequence of rotations  which will be illuminated shortly.

Setting aside this caveat until the next section, for the sake of simplicity we define $H_1$ to have the order of $a$-$b$ and $H^\prime_1$ have the order of $b$-$(a+b)$ where in this definition the first number is smaller (corresponding to the order of the largest off diagonal term).
It is easy to see that the rotation angle which can extinguish $(H^\prime_1)_{23}$ must be in order of $\mathcal{O}(\epsilon^b)$ and after that the rotated perturbative Hamiltonian must have order of $(a+b)$-$(a+b+b)$.
This is the famous Fibonacci sequence; that is, the order of the size of the largest off-diagonal element is the sum of the order after each of the previous two rotations.
The order of the smallness parameter in the perturbative part of the Hamiltonian will increase exponentially in the number of rotations since the Fibonacci sequence grows exponentially.
This means that the diagonal part of the Hamiltonian will converge on the true expression very rapidly.

A useful special case of $H_1$ is $a=b=1$.
That is, $H_1\propto\eps$.
We define $H^{(N)}_1$ to be the perturbative Hamiltonian after $N$ rotations (not including the preliminary rotations).
Then, we have that the size of the Hamiltonian shrinks exponentially as described by
\begin{equation}
\log_\epsilon H^{(N)}_1\sim-\frac{1}{\sqrt{5}}\left(\frac{1+\sqrt{5}}{2}\right)^N\,.
\end{equation}

Moreover, we notice that all the perturbative Hamiltonian's diagonal elements are zero.
Since the first order corrections to the eigenvalues are the diagonal elements of the perturbative Hamiltonian, therefore the order of errors of the eigenvalues will be double of the order of the perturbative Hamiltonian \cite{Denton:2019ovn}.

\begin{figure}[h]
\includegraphics[width=0.49\textwidth]{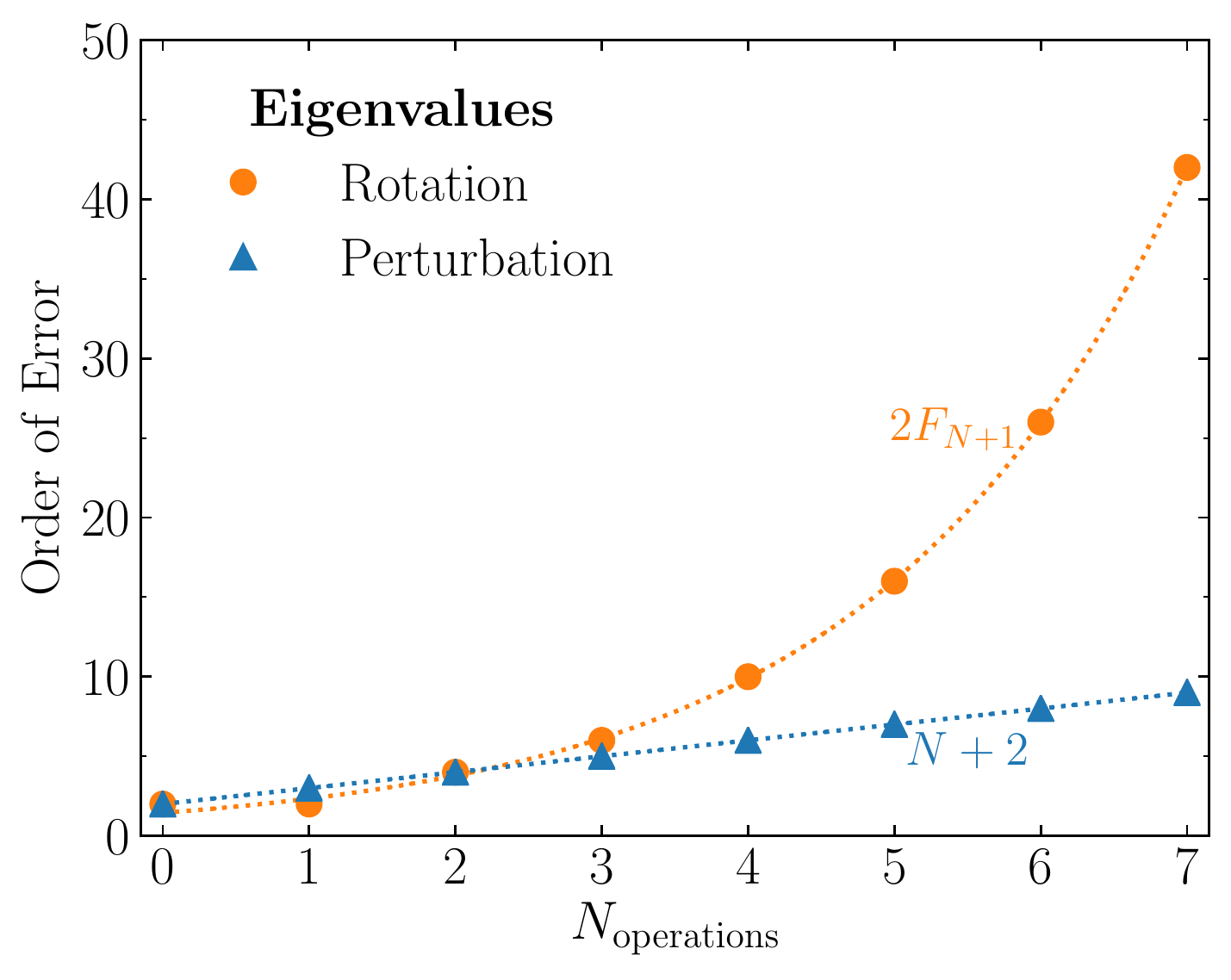}
\includegraphics[width=0.49\textwidth]{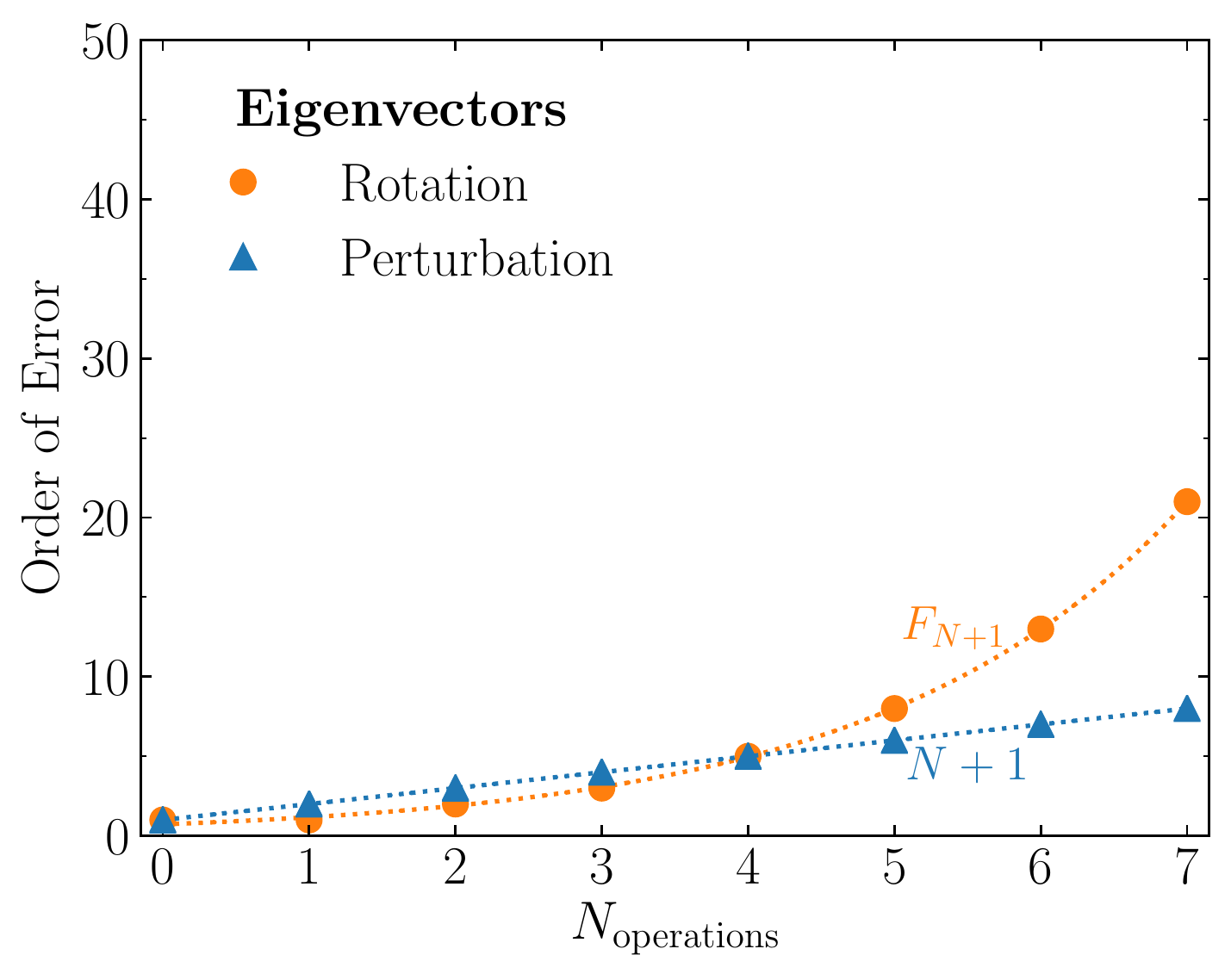}
\caption{Here we show the relative growth in precision by using rotations or following perturbation theory for $a=b=1$.
The horizontal axis is the number of operations: either the number of rotations of the order in perturbation theory.
The vertical axis shows the power $m$ of the size of the error, $\eps^m$.
{\bf Left:} The order of the error of the eigenvalues scales like $n+2$ using perturbation theory and $2F_{n+1}$ using rotations where $F_n$ is the $n^{\rm th}$ Fibonacci number.
{\bf Right:} The order of the error of the eigenvectors scales like $n+1$ using perturbation theory and $F_{n+1}$ using rotations.}
\label{fig:fib precision}
\end{figure}

For $a=b=1$, we compare orders of errors of the eigenvalues and eigenvectors given by perturbation expansions and the rotation method in Fig.~\ref{fig:fib precision}.
The order of the size of the error in the eigenvalues (eigenvectors) grows like $2F_{n+1}$ ($F_{n+1}$) where $F_n$ is the Fibonacci sequence defined as $F_0\equiv0$, $F_1\equiv1$, and $F_n=F_{n-1}+F_{n-2}$ for $n>1$.

\newpage

\section{Neutrino Oscillations in Matter}\label{section:numerical}
Now we illustrate this procedure in the context of neutrino oscillations in matter.
In flavor basis the Hamiltonian in matter is 
\begin{equation}
H=\frac{1}{2E}\left[U_\text{PMNS}\,\text{diag}(0,\Delta m^2_{21},\Delta m^2_{31})\,U^\dagger_\text{PMNS}+\text{diag}(a(x),0,0)\right]\,.
\end{equation}
$U_\text{PMNS}\equiv U_{23}(\theta_{23}, \delta)U_{13}(\theta_{13})U_{12}(\theta_{12})$ is known as the PMNS matrix. Note, this $U_\text{PMNS}$ is not the form usually used but it is equivalent after re-phasing $\nu_\tau$ and $\nu_3$, see \ref{App:pmns}. We use this form so that performing a vacuum rotation with $U_{23}(\theta_{23},\delta)$ gives us a real, symmetric Hamiltonian, a significant simplification.  With the usual form of the PMNS matrix the Hamiltonian is always complex if CP is violated, i.e. $\delta \neq 0$ or $\pi$.

  In the Earth's crust the matter potential is nearly constant \cite{Kelly:2018kmb,King:2020ydu},
\begin{align}
a(x)&\equiv2\sqrt{2}G_FN_eE \notag \\&\simeq 1.52\times 10^{-4}\Big(\frac{Y_e\rho}{\text{g}/\text{cm}^{3}}\Big)\Big(\frac{E}{\text{GeV}}\Big)~\text{eV}^2\,.
\end{align}
We perform the preliminary vacuum rotation on the (2-3) sector, since the matter potential commutes with this rotation, as follows:
\begin{equation}
\tilde{H}\equiv U^\dagger_{23}(\theta_{23},\delta)HU_{23}(\theta_{23},\delta)\,.
\label{eq:vac 23 rotation}
\end{equation}
Now $\tilde{H}$ is real and does not depend on $\theta_{23}$ and $\delta$; 
\begin{equation}
\tilde{H}=\tilde{H}_0+\tilde{H}_1\,,
\end{equation}
with
\begin{align}
\tilde{H}_0={}&\frac{1}{2E}~ \text{diag}(\lambda_a, \lambda_b, \lambda_c)
\quad  \text{where} \quad \begin{array}{l}
\lambda_a=a+(s^2_{13}+\epsilon s^2_{12})\Delta m^2_{ee}\,,\\[2mm]
\lambda_b=\epsilon c^2_{12}\Delta m^2_{ee}\,,  \\[2mm]
\lambda_c=(c^2_{13}+\epsilon s^2_{12})\Delta m^2_{ee}\,. 
\end{array}
\end{align}
The small scale $\epsilon=\Delta m^2_{21}/\Delta m^2_{ee}\sim0.03$, with $\Delta m^2_{ee}=c^2_{12}\Delta m^2_{31}+s^2_{12}\Delta m^2_{32}$ \cite{Nunokawa:2005nx}.
The off-diagonal part of the Hamiltonian is,
\begin{equation}
\tilde{H}_1=
\frac{\Delta m^2_{ee}}{2E}
\begin{pmatrix}
&\epsilon s_{12}c_{12}c_{13} & s_{13}c_{13} \\
\epsilon s_{12}c_{12}c_{13}& &-\epsilon s_{12}c_{12}s_{13} \\
 s_{13}c_{13} &-\epsilon s_{12}c_{12}s_{13} & \\
\end{pmatrix}\,.
\label{eq:H1}
\end{equation}
There are two important things to note about the form of the exact Hamiltonian in this basis:
\begin{itemize}
\item The crossings of the diagonal elements: $\lambda_a=\lambda_c$ occurs when  $a=\Delta m^2_{ee} \cos 2 \theta_{13}$, i.e.~at the atmospheric resonance. Whereas $\lambda_a=\lambda_b$ occurs when $a=\Delta m^2_{21} \cos 2 \theta_{12}-\Delta m^2_{ee} s^2_{13}$. This, however, is significantly different from  $a= \Delta m^2_{21} \cos 2 \theta_{12}/c^2_{13}$, the solar resonance.
Since 
$\Delta m^2_{ee} s^2_{13}> \Delta m^2_{21} \cos 2 \theta_{21}$,  for normal ordering, $\lambda_a=\lambda_b$  occurs when $a<0$, i.e.~for anti-neutrinos not neutrinos.
There is no crossing between $\lambda_b$ and  $\lambda_c$.
\item There is a significant hierarchy (approximately an order of magnitude) in the off diagonal elements, since  $|\tilde{H}_{13}| \gg |\tilde{H}_{12}| \gg |\tilde{H}_{23}|$, since 
\begin{align}
s_{13}c_{13} \approx 0.15, \quad \epsilon s_{12}c_{12}c_{13} \approx 0.015 \quad \text{and} \quad   \epsilon s_{12}c_{12}s_{13} \approx 0.0021.
\nonumber
\end{align}
And the off-diagonal elements are independent of the matter potential.
\end{itemize}
These two features have significant impact on how effective the LODE or LROT strategy has on reducing the off-diagonal elements by additional rotations.  

\subsection{LODE strategy}
After the vacuum $U_{23}$ rotation, if we apply the LODE strategy, i.e.~perform a (1-3) rotation to set the $(H_1)_{13}=(H_1)_{31}=0$, we have
\begin{align}
\hat{H}\equiv{}&U^\dagger_{13}(\tilde{\theta}_{13})\tilde{H}U_{13}(\tilde{\theta}_{13}) \notag
= \frac{1}{2E}
\begin{pmatrix}
\lambda_-& & \\
&\lambda_0 & \\
& &\lambda_+ \\
\end{pmatrix}
\notag\\[2mm]
&
+\epsilon c_{12}s_{12}\frac{\Delta m^2_{ee}}{2E}
\begin{pmatrix}
&\cos(\tilde{\theta}_{13}-\theta_{13}) & \\
\cos(\tilde{\theta}_{13}-\theta_{13})& & \sin(\tilde{\theta}_{13}-\theta_{13})\\
&\sin(\tilde{\theta}_{13}-\theta_{13}) & \\
\end{pmatrix}\,,
\label{eq:Hlode}
\end{align}
where
\begin{align}
\lambda_\pm={}&\frac{1}{2}\left[(\lambda_a+\lambda_c)\right.\notag\\
&\left.\pm\text{sign}(\Delta m^2_{ee})\sqrt{(\lambda_a-\lambda_c)^2+4(s_{13}c_{13}\Delta m^2_{ee})^2}\right]\,, \notag \\
\lambda_0={}&\epsilon c^2_{12}\Delta m^2_{ee}\,.
\end{align}
With the diagonal elements above, $\tilde{\theta}_{13}$ can be determined by
\begin{equation}
\sin^2\tilde{\theta}_{13}=\frac{\lambda_+-\lambda_c}{\lambda_+-\lambda_-}
=\frac12\left(1-\frac{\lambda_c-\lambda_a}{\lambda_+-\lambda_-} \right)
\,, \quad \tilde{\theta}_{13}\in [0,\pi/2]\,.
\end{equation}
For more details see \ref{App:rots}.

Now consider $a \approx \Delta m^2_{21}$ then
\begin{align}
\lambda_- = {}& c^2_{13} a + s^2_{12} \Delta m^2_{21} +{\cal O}(a^2/\Delta m^2_{ee}) \, , \notag \\
\lambda_0={}&\lambda_b	= c^2_{12}\Delta m^2_{21}\, , \notag \\
\lambda_+= {}& s^2_{13} a + \Delta m^2_{31} +  {\cal O}(a^2/\Delta m^2_{ee})\,.\
\end{align}
Now $\lambda_-=\lambda_0$ occurs at the solar resonance, $a= \Delta m^2_{21} \cos 2 \theta_{12}/c^2_{13}$. Also, we have
\begin{align}
 \sin(\tilde{\theta}_{13}-\theta_{13})= s_{13} c_{13} ( a/\Delta m^2_{ee}) \{1+{\cal O}[ a/\Delta m^2_{ee}]\}\,,
\end{align}
which at the solar resonance is  $\sim s_{13} \epsilon$. Therefore, if we now perform a (1-2) rotation, there is a cancellation between numerator and denominator in the correction terms near the solar resonance.  Exactly at the point you might expect that the LODE strategy would perform poorly. 
For larger $a$, when $\tilde{\theta}_{13} > \pi/4+\theta_{13}$, the LODE strategy switches to performing a (2-3) rotation next, but for such values of the matter potential all of the difference of the diagonal elements are ${\cal O}(\Delta m^2_{ee})$ or larger.

\subsection{LROT strategy}

If, after the vacuum $U_{23}$ rotation, we apply the LROT strategy for an $a \approx \Delta m^2_{21}$, i.e.~perform a (1-2) rotation to set the $(H_1)_{12}=(H_1)_{21}=0$, we have
\begin{align}
\hat{H}\equiv{}&U^\dagger_{12}(\tilde{\theta}_{12})\tilde{H}U_{12}(\tilde{\theta}_{12}) \notag
= \frac{1}{2E}
\begin{pmatrix}
\lambda_\rho& & \\
&\lambda_\sigma & \\
& &\lambda_\tau \\
\end{pmatrix}
\notag\\[2mm]
&    \hspace{-5mm}
+s_{13}\sqrt{c^2_{13}+\epsilon^2 s^2_{12} c^2_{12}} \left( \frac{\Delta m^2_{ee}}{2E} \right)
\begin{pmatrix}
& & \cos(\tilde{\theta}_{12}-\omega)  \\
& & \sin(\tilde{\theta}_{12}-\omega)\\
\cos(\tilde{\theta}_{12}-\omega) &\sin(\tilde{\theta}_{12}-\omega) & \\
\end{pmatrix}\,,
\label{eq:Hrot}
\end{align}
where
\begin{align}
\lambda_{\rho,\sigma}={}&\frac{1}{2}\left[(\lambda_a+\lambda_b)
\pm\sqrt{(\lambda_a-\lambda_b)^2+4(c_{13}s_{12}c_{12}\Delta m^2_{21})^2}\right]\,, \notag \\
\lambda_\tau={}&\lambda_{c}= (c^2_{13}+\epsilon s^2_{12})\Delta m^2_{ee}\,. 
\label{eq:rho}
\end{align}
With the diagonal elements above, $\tilde{\theta}_{12}$ can be determined by
\begin{equation}
\sin^2\tilde{\theta}_{12}=\frac{\lambda_\sigma-\lambda_b}{\lambda_\sigma-\lambda_\rho}
=\frac12\left(1-\frac{\lambda_b-\lambda_a}{\lambda_\sigma-\lambda_\rho} \right)
\,, \quad \tilde{\theta}_{12}\in [0,\pi/2]\,.
\end{equation}
The angle $\omega$ is given by
\begin{align}
 \cos \omega= & c_{13}\biggr/\sqrt{c^2_{13}+\epsilon^2 s^2_{12} c^2_{12}}, \quad 
 \sin \omega=  \epsilon s_{12} c_{12}\biggr/\sqrt{c^2_{13}+\epsilon^2 s^2_{12} c^2_{12}},
\end{align}
therefore $\omega \approx \epsilon  s_{12} c_{12} \ll 1$.

Now the ``resonance'' in Eq.~\ref{eq:rho}  occurs when $\lambda_a=\lambda_b$, which is  $a = \cos 2\theta_{12}\Delta m^2_{21}-s^2_{13} \Delta m^2_{ee}$ which is significantly far  from the solar resonance $a= \Delta m^2_{21} \cos 2 \theta_{12}/c^2_{13}$.  This is the reason why the LROT strategy is a poor one near the solar resonance, exactly where one might expect it to be superior to the LODE strategy.

After the vacuum (2-3) rotation and the matter (1-3), the off-diagonal part of the Hamiltonian, see eq.\ref{eq:Hlode}, is of order
$$ s_{12} c_{12} \Delta m^2_{21}/(2E), $$
whereas after the vacuum (2-3) rotation and then matter (1-2) rotation, the off-diagonal part of the Hamiltonian, eq.\ref{eq:Hrot}, is of order
$$ s_{13} c_{13} \Delta m^2_{ee}/(2E)\, .$$
Note that for the known neutrino parameters, the first of these is significantly smaller than the latter.  So that the LODE strategy has diminished the over all size the off-diagonal part of the Hamiltonian more than the LROT strategy for all values of the matter potential.
This effect continues in subsequent rotations and has significant accumulative effects on the size of the off-diagonal part of the Hamiltonian.
 If we consider a scenario such that $s_{13}$ is an order of magnitude  smaller so that $ s_{12} c_{12} \Delta m^2_{21} > s_{13} c_{13} \Delta m^2_{ee}\, .$  In this scenario, the vacuum (2-3) plus matter (1-2) reduces the off-diagonal part of the Hamiltonian more than vacuum (2-3) plus matter (1-3).  However, since this change in the size of $s_{13}$ also changes the relative size of the elements of $H_1$, eq. \ref{eq:H1}. Therefore it also changes the rotation after the vacuum (2-3) rotation, in the LODE strategy.   So that even with this significant change in $s_{13}$ the LODE strategy also adjusts to accommodate this change. We have checked numerically that in this scenario that LODE strategy works better than LROT especially around the solar resonance.

To summarize here we have found  the surprising result that at least for three flavor neutrino oscillations in matter the LODE strategy is superior or equal to  LROT strategy for all values of the matter potential.
It is exactly in the resonance regions that the LODE strategy is superior to LROT strategy as noted above and will be confirmed numerically for all rotations in the next subsection.

\subsection{Numerical Comparison of LODE versus LROT}
\label{sec:num}

The metric we will use for the size of the corrections after each rotation is the size of the first order corrections to the eigenvectors after this rotation\footnote{The first order corrections to eigenvalues vanish as $H_1$ has zeros on the diagonal.}, given by
\begin{equation}
\text{max}_{\, j>k} \left| 2E\,(H_1)_{jk} /\Delta \lambda_{jk} \right|\,.  \label{eq:metric0}
\end{equation}
Note that this metric gives larger corrections near the resonances as expected in these regions\footnote{After the first rotation the difference between this metric and  using the metric   $\text{max}_{\, j>k} | 2E\, (H_1)_{jk}   /\Delta m^2_{ee} | $ is small.}.  The vacuum parameters used are $\sin^2\theta_{12}=0.31$, $\sin^2\theta_{13}=0.022$, $\sin^2\theta_{23}=0.58$, $\delta_{CP}=215^\circ$, $\Delta m^2_{21}=7.4\times 10^{-5}$ eV$^2$, and $\Delta m^2_{31}=2.5\times 10^{-3}$ eV$^2$ \cite{Esteban:2018azc}.  Positive $Y_e \rho E$ corresponds to neutrinos while negative is anti-neutrinos.  The figures below are for normal ordering.

\begin{figure}[t]
\centering
\includegraphics[width=0.48\textwidth]{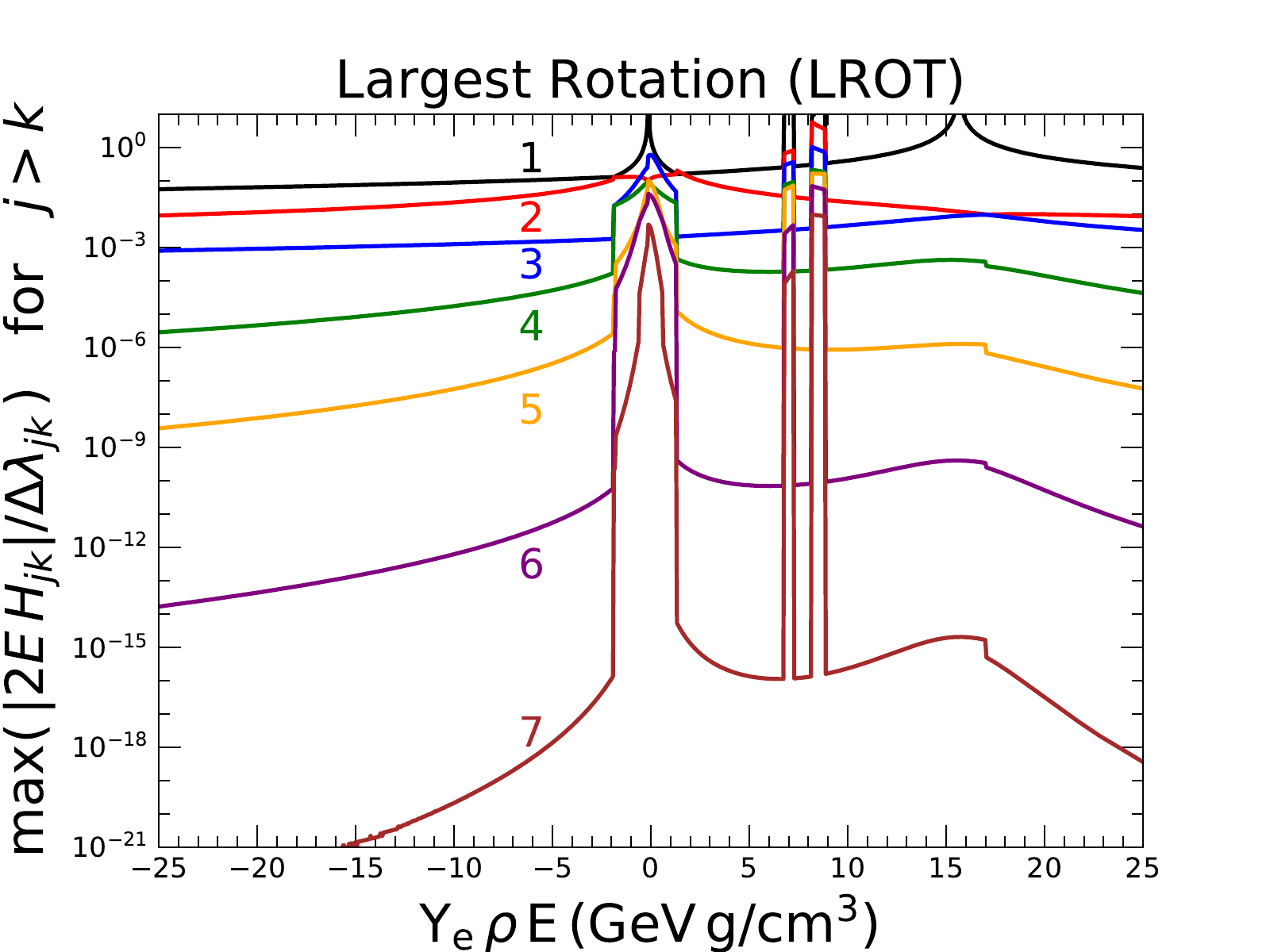}
\includegraphics[width=0.48\textwidth]{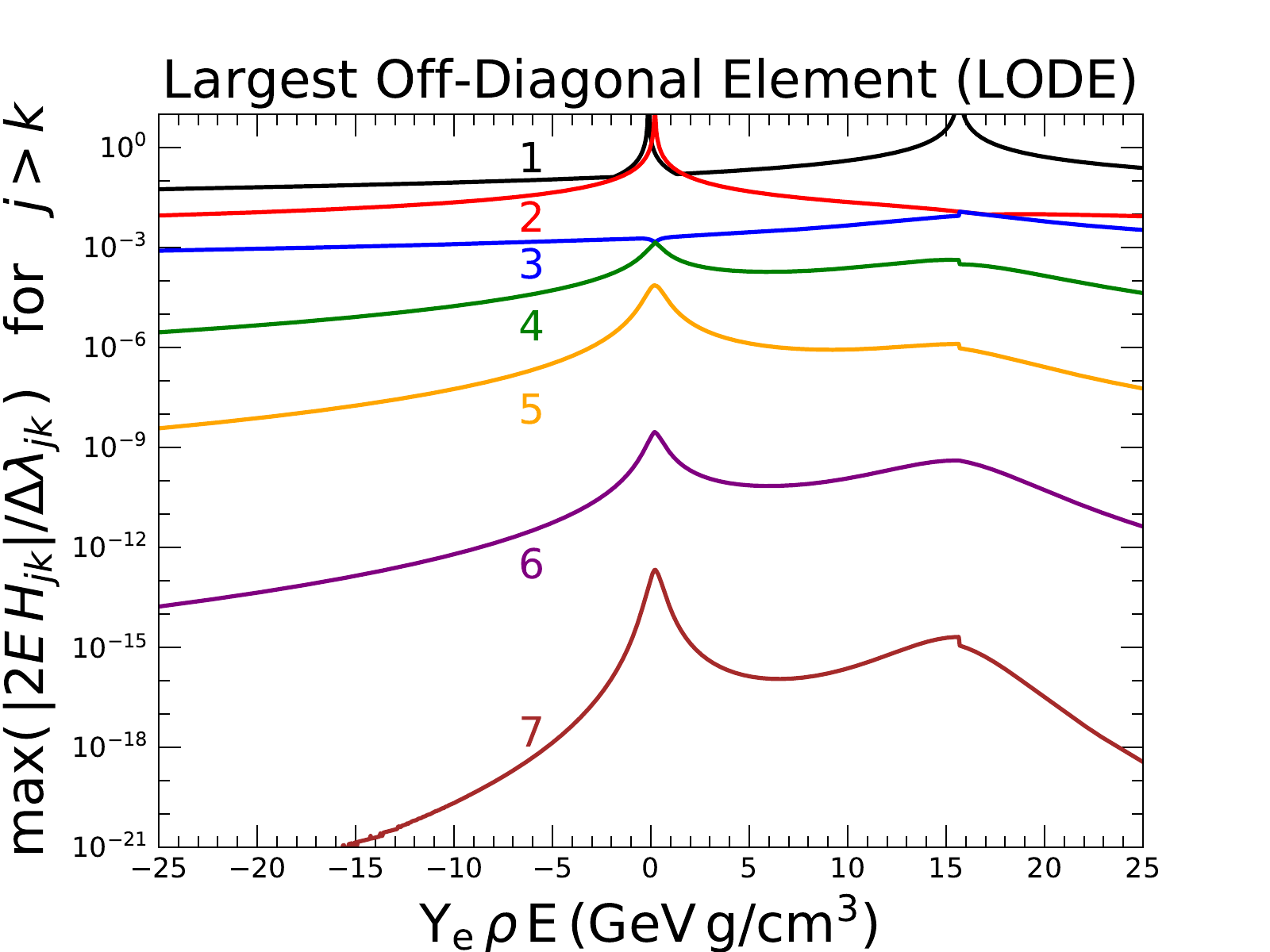}\\
\includegraphics[width=0.48\textwidth]{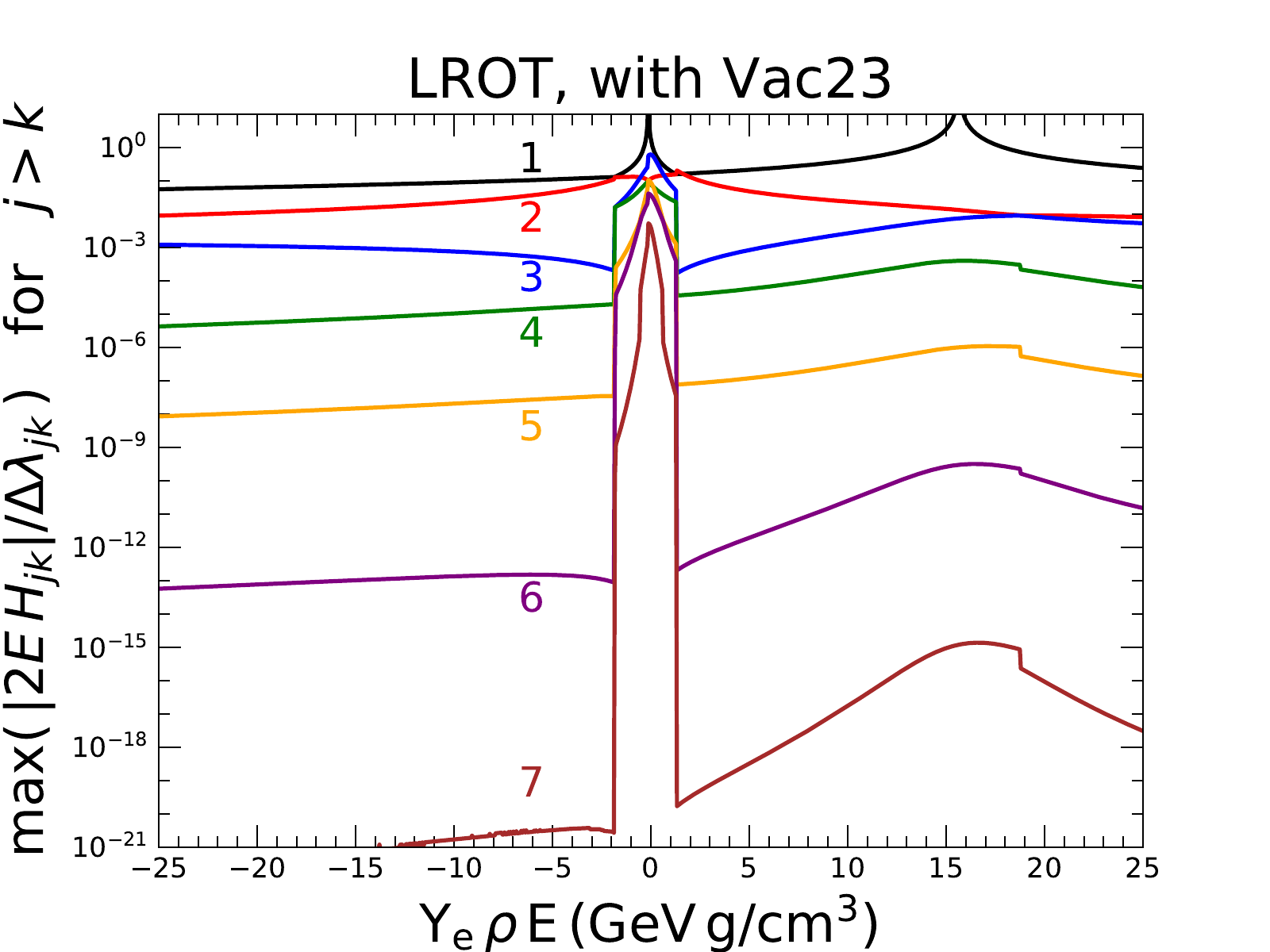}
\includegraphics[width=0.48\textwidth]{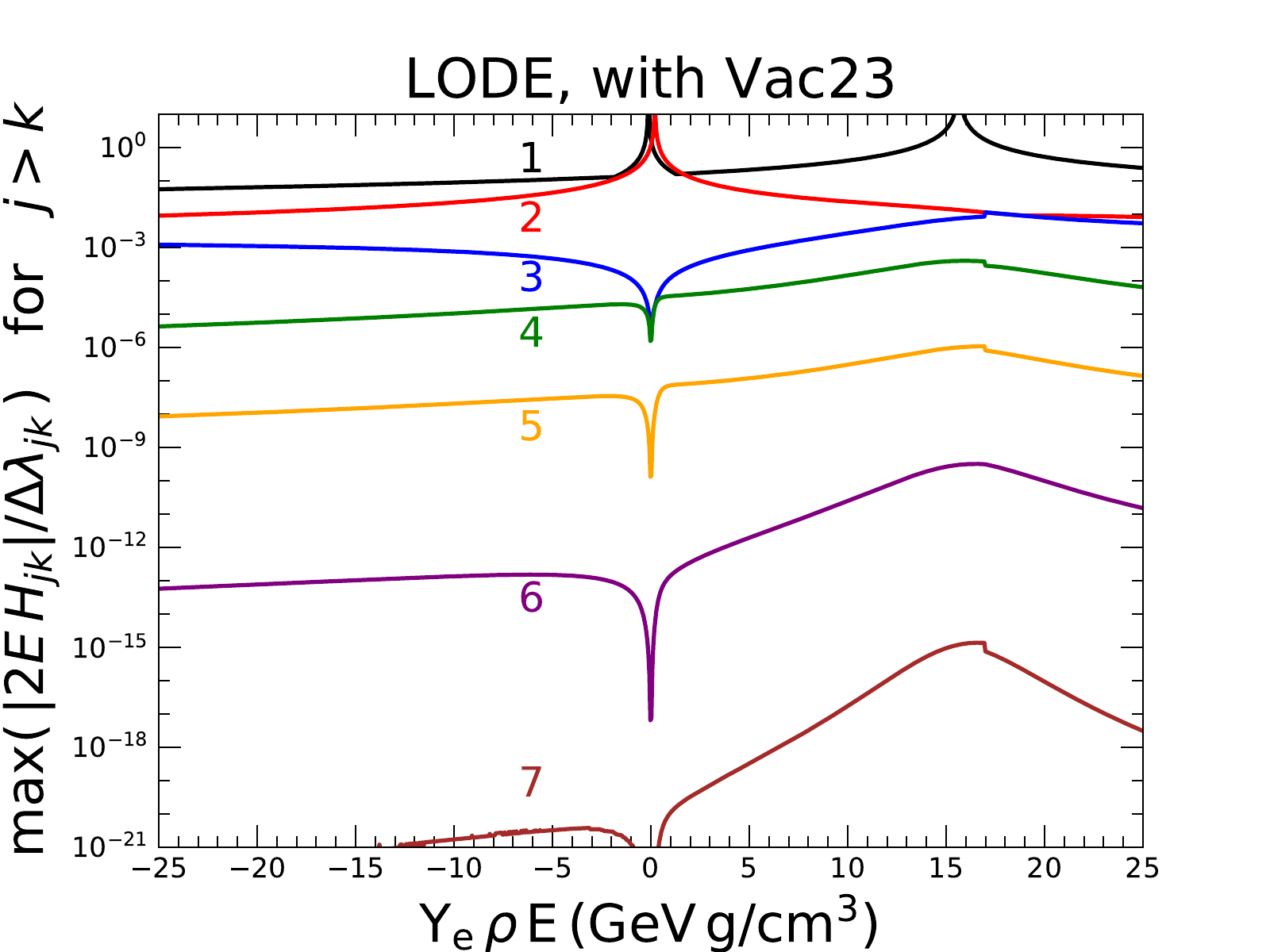}
\caption{The size of the corrections to the eigenvectors, $\text{max}_{\, j>k} \left| 2E\,(H_1)_{jk} /\Delta \lambda_{jk} \right|$
after $N$ rotations including any initial rotations  as a function of the neutrino energy times 
$Y_e \rho$.
{\bf Left (Right) panels}: The pivot is selected using the LROT (LODE) strategy.
{\bf Top  (Bottom) row}: Without (With) an initial vacuum (2-3) rotation.
The LROT strategy has convergence issues wherever the difference in the diagonal elements becomes small. }
\label{fig:H1_before23}
\end{figure}

In Fig.~\ref{fig:H1_before23}, top row, we display the size of the first order corrections after $N$ rotations  for both the LROT and LODE strategy {\it without} a vacuum (2-3) rotation to start. Note that at the level crossings of the diagonal elements  the LROT strategy does poorly.  The LODE strategy does reasonably well every where but note that even in vacuum, $a=0$, the  corrections to eigenvectors are not zero. This occurs because the (2-3) rotation here is not exactly the vacuum rotation because of the tiny $(H_1)_{23}$ term in Eq.~\ref{eq:H1}.

In Fig.~\ref{fig:H1_before23}, bottom row, we display the size of the first order corrections after $N$ rotations  for both the LODE and LROT strategy {\it  with} a vacuum (2-3) rotation first. Now the LROT strategy does well except around the solar resonance for the reasons discussed earlier. The LODE strategy does very well everywhere and the corrections follow the Fibonacci sequence even at the solar resonance due to the cancellation between numerator and  denominator of Eq.~\ref{eq:metric0} as discussed earlier.

If one  imposes both a vacuum (2-3) rotation  and a full (1-3) rotation, both the LROT and LODE strategies give very similar results.
Imposing an additional (1-2) rotation after the vacuum (2-3) and a full (1-3) rotation also gives similar results but above the atmospheric resonance a better choice for the third rotation is (2-3) rather than (1-2).
All of these statements are summarized in the Table \ref{tab:summary}.
In addition, \emph{only} an initial vacuum (2-3) rotation makes the resulting Hamiltonian real unless CP happens to be conserved.
That is, we find that for neutrino oscillations in matter performing the vacuum (2-3) rotation first is the optimal scenario, which isn't too surprising since the (2-3) vacuum rotation results in a real matrix, leaves the matter potential unchanged, and somewhat simplifies the other part.  Also be performing the vacuum (2-3) rotation instead of the (2-3) rotation that sets  $H_{23}=0$, one insures that in the LODE strategy the corrections are zero in vacuum, after two additional rotations. This can be seen in the right hand panels of  Fig. \ref{fig:H1_before23}. 

Naively, one might have expected in the region around the solar resonance, where there is a hierarchy in the differences of the diagonal elements of the Hamiltonian, that LROT strategy would be better than LODE strategy. This is clearly {\it not} the case and the reason is that without the (1-3) rotation first, the crossing of the diagonal elements of the Hamiltonian does not occur at the solar resonance as was noted in the comments after Eq.~\ref{eq:H1}.  In Table \ref{tab:one}, we have given the sequence of rotations for an energy just above the solar resonance.  The LROT strategy always chooses a (1-2) rotation unless the previous rotation is a (1-2) rotation which has set the $(H_1)_{12}=0$.  Near the solar resonance the LROT preference for the (1-2) rotation is too strong for rapid convergence.  However, as expected, the order spans of each rotation increase with the total number of rotations for both LODE and LROT strategies but the LROT strategy converges considerably more slowly around the solar resonance.

\begin{table}[h]
\centering
\begin{tabular}{|cc|cccc|}
\cline{3-6}
\multicolumn{2}{c|}{}&\multicolumn{4}{c|}{Initial rotation(s)}\\
\multicolumn{2}{c|}{}&None&(2-3)$_\text{v}$&(2-3)$_\text{v}$+(1-3)&(2-3)$_\text{v}$+(1-3)+(1-2) \\[1mm]
\hline
\multirow{2}{*}{Strategy}&LROT&\xmark \xmark & \xmark & \cmark\cmark\cmark & \cmark \cmark \\[1mm]
&LODE&\cmark & \cmark \cmark \cmark &\cmark \cmark \cmark &\cmark \cmark \\
\hline
\end{tabular}
\caption{The effect of the different strategies on the convergence rate of the precision are ranked, after the initial rotation(s), from None to Vacuum (2-3)+Matter (1-3)+Matter (1-2), for both the LROT and LODE strategies for all neutrino energies:  from poor (\xmark \xmark) to excellent (\cmark \cmark \cmark). Only the (2-3)$_\text{v}$ is a vacuum rotation; every other rotation diagonalizes the indicated $2\times2$ submatrix of the Hamiltonian. The three cases with  \cmark \cmark \cmark's have approximately equal convergence for all energies.  Beyond the atmospheric resonance the two cases with  \cmark \cmark's have slightly slower convergence than those with   \cmark \cmark \cmark's. }
\label{tab:summary}
\end{table}

\begin{table}[h]
\begin{center}
\vspace*{4mm}
\begin{tabular}{|cc|ccccccc|}
\cline{3-9}
\multicolumn{2}{c|}{}&\multicolumn{7}{c|}{Rotation~\#}\\
\multicolumn{2}{c|}{}& 1 & 2& 3& 4& 5& 6&7 \\
\hline
\multirow{2}{*}{Strategy}&LROT&(2-3)$_\text{v}$& (1-2) & (2-3) &(1-2) & (1-3) & (1-2) & (2-3) \\[1mm]
 &LODE&(2-3)$_\text{v}$& (1-3) & (1-2) & (1-3) & (2-3)$^\dagger$ & (1-2) & (1-3) \\
\hline
\end{tabular}
\caption{The sequence of matter rotation in both the LROT and LODE strategies with a (2-3)$_\text{v}$ rotation first (the second column in Table \ref{tab:summary}).
We consider the case of a matter potential given by $Y_e \rho E= 0.250$ GeV g/cm$^3$, slightly above the solar resonance.
The (2-3)$_\text{v}$ is a vacuum rotation.
After the above sequences, the size of the corrections to the eigenvectors are $10^{-21}$ and $10^{-3}$ for LODE and LROT strategies, respectively.
$^\dagger$The LODE strategy achieves better than $10^{-8}$ after 4 matter rotations at this value of the matter potential. }
\label{tab:one}
\end{center}
\end{table}

The method could be simply extended to include neutrino non-standard interactions (NSIs) \cite{Wolfenstein:1977ue} and/or the addition of sterile neutrinos and a similar rate of convergence is expected.

\section{Conclusion}\label{section:conclusion}
We show the potential application of a rotation method in high precision calculations of neutrinos oscillations in matter.
The fast iteration steps of the method can enhance the zeroth order precision very rapidly.
More specifically, a Fibonacci recursive process leads to an exponential growth of the orders (of some small scale) of the zeroth order eigensystem's errors with number of the rotations.
This feature grants an advantage of the rotation method in the range of high precision calculation compared with the perturbation expansion methods.
We find that the method of selecting the pivot at the largest off-diagonal element performs best; selecting the pivot that leads to the largest rotation is similar in many cases but leads to convergence issues especially around the  solar resonance.
In addition, while the complexity of each additional step in a perturbative expansion grows, the complexity of each additional rotation is constant and simple as shown in Eqs.~\ref{eq:u}-\ref{eq:rotationangle}.


\section*{Acknowledgements}
We thank Serguey Petcov for helpful comments.
PBD acknowledges the United States Department of Energy under Grant Contract desc0012704 and the Neutrino Physics Center.
This manuscript has been authored by Fermi Research Alliance, LLC under Contract No.~DE-AC02-07CH11359 with the U.S.~Department of Energy, Office of Science, Office of High Energy Physics.
SP received funding/support from the European Unions Horizon 2020 research and innovation programme under the Marie Sklodowska-Curie grant agreement No 690575 and No 674896.

\appendix

\section{PMNS  Matrix}
\label{App:pmns}

\newcommand{\U}{U_{\rm PMNS}}
$\U$ is the lepton mixing matrix in vacuum, given by\\
$\U\equiv U_{23}(\theta_{23},\delta)U_{13}(\theta_{13})U_{12}(\theta_{12})$ with
\begin{equation}
\begin{gathered}
U_{12}(\theta_{12}) \equiv
\begin{pmatrix}
c_{12}&s_{12}\\
-s_{12}&c_{12}\\
&&1
\end{pmatrix}\,,\quad
U_{13}(\theta_{13}) \equiv
\begin{pmatrix}
c_{13}&&s_{13}\\
&1\\
-s_{13}&&c_{13}
\end{pmatrix}\,,\\[2mm]
\hspace*{-3mm}
U_{23}(\theta_{23}, \delta)\equiv
\begin{pmatrix}
1\\
&c_{23}&s_{23}e^{i\delta}\\
&-s_{23}e^{-i\delta}&c_{23}
\end{pmatrix} \, .
\end{gathered}
\label{eq:PMNS}
\end{equation}
The PDG form of $\U$ is obtained from our
$\U$ by multiplying the 3rd row by $e^{i\delta}$ and the 3rd column by $e^{-i\delta}$ i.e.~by rephasing $\nu_\tau$ and $\nu_3$.
The shorthand notation $c_\theta = \cos \theta$ and $s_\theta = \sin \theta$ is used throughout this paper.
This form is used such that after performing the vacuum (2-3) rotation, the resultant Hamiltonian is real symmetric, as shown in eq.~\ref{eq:vac 23 rotation}.

\section{Rotations of a 3x3 Symmetric  Matrix}
\label{App:rots}

Rotation to set $H_{13}=H_{31}=0$:
\begin{align}
&U^\dagger_{13}(\phi)
\begin{pmatrix}
\lambda_a&  \lambda_y &  \lambda_x\\
\lambda_y&\lambda_b & \lambda_z\\
\lambda_x& \lambda_z&\lambda_c \\
\end{pmatrix}
U_{13}(\phi) 
= \begin{pmatrix}
\lambda_\rho&  &   \\
&\lambda_\sigma & \\
& &\lambda_\tau \\
\end{pmatrix}
\notag \\
& +\sqrt{\lambda_y^2 +\lambda^2_z}
\begin{pmatrix}
& \cos(\phi-\omega) &   \\
 \cos(\phi-\omega)   &  & \sin(\phi-\omega) \\
& \sin(\phi-\omega) & \\
\end{pmatrix}
\end{align}
with $\sin^2 \phi=  (\lambda_\tau-\lambda_c)/(\lambda_\tau-\lambda_\rho)= \frac12 \left(1- \frac{\lambda_c-\lambda_a}{ \lambda_\tau-\lambda_\rho} \right)$  and
\begin{align}
\lambda_{\rho,\tau} =  &     \frac12 \left(\lambda_a+\lambda_c \pm \sqrt{ (\lambda_a-\lambda_c)^2+4\lambda^2_x}  ~\right), \quad
  \lambda_\sigma =     \lambda_b \,  , \notag \\
\cos \omega = & \lambda_y / \sqrt{\lambda_y^2 +\lambda^2_z} \quad \text{and} \quad 
\sin \omega =- \lambda_z / \sqrt{\lambda_y^2 +\lambda^2_z}.
\end{align}
 Note that the functional form of the rotated (12) and (23) elements is independent of the explicit value of $\phi$.

Similar results to set $H_{12}$ or $H_{23}$ to zero can be obtained by permuting these results.\\

\bibliographystyle{elsarticle-num}
\bibliography{Fibonacci}

\end{document}